\documentstyle[12pt,fleqn]{article}

\newcommand\lapprox{\mathrel{\mathop
  {\hbox{\lower0.5ex\hbox{$\sim$}\kern-0.8em\lower-0.7ex\hbox{$<$}}}}}
\newcommand\gapprox{\mathrel{\mathop
  {\hbox{\lower0.5ex\hbox{$\sim$}\kern-0.8em\lower-0.7ex\hbox{$>$}}}}}

\begin{document}

\title{Pulsar Bound on the Photon Electric Charge Reexamined}

\author{Georg Raffelt\thanks{{\it Permanent Address:\/}
Max-Planck-Institut f\"ur Physik, F\"ohringer Ring 6, 80805 M\"unchen,
Germany}\\
Center for Particle Astrophysics, University of California\\
Berkeley, CA 94720, U.S.A.}

\maketitle

\begin{abstract}
\noindent If photons had a small electric charge $Q_\gamma$ their path
in the galactic magnetic field would be curved, leading to a time
delay between photons of different frequency from a distant source.
Cocconi's previous application of this argument led to a limit which
is too restrictive by a factor of about 200; the corrected bound is
$Q_\gamma/e\lapprox10^{-29}$.\par
%\bigskip
%\noindent {\it PACS categories:} 14.70.Bh, 97.60.Gb;
\end{abstract}

%%%%%%%%%%%%%%%%%%%%%%%%%%%%%%%%%%%%%%%%%%%%%%%%%%%%%%%%%%%%%%%%%%%%%
\bigskip
%\newpage

\noindent Astronomical time-of-flight methods allow one to set
powerful limits on hypothetical electric charges of neutrinos and
photons.  Their path in the galactic magnetic field would be curved,
leading to an energy-dependent time delay. The absence of an anomalous
spread of the neutrino arrival times from SN~1987A allowed Barbiellini
and Cocconi \cite{Barbiellini} (see also Bahcall \cite{Bahcall}) to
derive a very restrictive limit of
$Q_{\nu_e}/e\lapprox3{\times}10^{-17}$. The same method applied to the
radio pulses from the pulsar PSR 1937+21 led Cocconi \cite{CocconiI}
to claim a bound on a putative photon electric charge of
$Q_\gamma/e\lapprox2{\times}10^{-32}$, a result which is the
``standard limit'' quoted in the summary tables of the Review of
Particle Properties (see \cite{ParticleData} for the most recent
edition). Unfortunately, this bound is not correct; it must be relaxed
by an approximate factor 200.

Relativistic particles (energy $E$) emitted from a source at a
distance $\ell$ take the time $t=\ell$ to reach Earth (natural units
with $\hbar=c=1$ are used). If these particles have a small charge $Q$
and move in a homogenous transverse magnetic field $B$ their path is
curved whence they are delayed by~\cite{Barbiellini,Bahcall}
\begin{equation}\label{001}
\frac{\Delta t}{t}=\frac{Q^2 B^2\ell^2}{6 E^2}.
\end{equation}
If these particles had a small mass they would be delayed by
\begin{equation}\label{002}
\frac{\Delta t}{t}=\frac{m^2}{2E^2}
\end{equation}
with the same $E^{-2}$ scaling. Therefore, the absence of a frequency
dependent dispersion of a short pulse can be used to limit a particle
mass and its charge in the same way; this has been done for the
neutrinos from SN~1987A~\cite{Bahcall}.

For photons, the standard dispersion effect from the interstellar gas
must be included. In a nonrelativistic ionized medium the photon
dispersion relation is such that it mimics a photon mass
$m_\gamma=\omega_{\rm P}$ (plasma frequency) given by
$\omega_{\rm P}^2=4\pi\alpha\,n_e/m_e$ (fine-structure constant
$\alpha$, electron density $n_e$, electron mass $m_e$).  Thus, the
standard photon delay is given by Eq.~(\ref{002}) with
$m\to\omega_{\rm P}$; typical interstellar electron densities are of
order $0.1\,\rm cm^{-3}$, corresponding to $\omega_{\rm P}$ of order
$10^{-11}\,\rm eV$.  Therefore, it is well known that the observed
dispersion of pulsar signals yields limits on a vacuum photon mass not
better than the $10^{-11}\,\rm eV$ level. Cocconi \cite{CocconiI}
remarked on this problem but failed to acknowledge that it also
persists for the case of a photon charge because the magnetic field
delay of Eq.~(\ref{001}) has the same $E^{-2}$ scaling as the standard
dispersion effect.  Therefore, the entire observed signal dispersion
could be blamed on a photon charge whereas Cocconi only allowed the
observational uncertainty of the dispersion measure as the maximum
effect caused by $Q_\gamma$.

In order to derive the corrected limit one may write the observed
dispersion effect in the form $\Delta t=D/f^2=(2\pi)^2D/E^2$ where
$E=h f=2\pi\hbar f=2\pi f$ was used in natural units (frequency $f$,
dispersion constant $D$).  If the entire observed $D$ is caused by the
charge effect of Eq.~(\ref{001}) one finds
\begin{equation}\label{003}
\frac{Q_\gamma}{e}=
\left(\frac{6\pi D}{\alpha B^2 \ell^3}\right)^{1/2}
=1.08{\times}10^{-29}\,\frac{1\,\mu{\rm G}}{B}\,
\left(\frac{D}{10^{17}\,\rm s^{-1}}\right)^{1/2}\,
\left(\frac{1\,\rm kpc}{\ell}\right)^{3/2}.
\end{equation}
(For the conversion of units note that $1\,\rm G$ corresponds to
$1.95{\times}10^{-2}\,\rm eV^2$ in natural, rationalized units where
$\alpha=e^2/4\pi$.) For the pulsar PSR~1937+21 one finds
$D=(29.479\pm0.001)\times10^{16}\,\rm s^{-1}$ \cite{Rawley}. A typical
galactic transverse $B$ is $1\mu\rm G$, and $\ell>2.5\,\rm kpc$
\cite{Rawley}. Then
\begin{equation}\label{004}
\frac{Q_\gamma}{e}\lapprox5{\times}10^{-30},
\end{equation}
about a factor of 200 less restrictive than Cocconi's original result.

I thank G.~Cocconi for a personal communication in which he expressed
agreement with my re-analysis of his original argument, and for
calling my attention to a different method for deriving a limit on the
photon charge of about $Q_\gamma/e\lapprox10^{-28}$ \cite{CocconiII}.

%%%%%%%%%%%%%%%%%%%%%%%%%%%%%%%%%%%%%%%%%%%%%%%%%%%%%%%%%%%%%%%%%%%%%%

%\newpage

\end{document}